%% file: main.tex
\documentclass[journal]{IEEEtran}

\input{Macro}

\input{figs/plotstyle}
\usepackage{eso-pic}
\AddToShipoutPictureFG*{%
  \AtPageUpperLeft{%
    \raisebox{-2.2\baselineskip}{%
      \makebox[\paperwidth]{%
        \parbox{0.82\paperwidth}{\centering\footnotesize
          This work has been submitted to the IEEE for possible publication.
          Copyright may be transferred without notice, after which this
          version may no longer be accessible.}}}}}

\begin{document}
\bstctlcite{IEEEexample:BSTcontrol}

\title{Split-Ring Couplers for On-Skin Wi-Fi Communication via an SSPP Waveguide Textile}

\author{Akihito~Noda,~\IEEEmembership{Member,~IEEE}%
~and~Ryo~Takahashi,~\IEEEmembership{Member,~IEEE}%
\thanks{A. Noda is with the School of Systems Engineering, Kochi University of Technology, Kochi 782-8502, Japan e-mail: (see https://orcid.org/0000-0002-6393-3196).}
\thanks{R. Takahashi is with the Department of Electrical Engineering and Information Systems, The University of Tokyo, Tokyo 113-8656, Japan e-mail: (see https://orcid.org/0000-0001-5045-341X)}%
\thanks{This research was supported in part by
JST, CRONOS, Japan Grant Number JPMJCS25N4,
and JSPS KAKENHI Grant Numbers JP25K03109,
JP22K21343.}%
\thanks{Corresponding author: Akihito Noda}%
}

\maketitle

\begin{abstract}
Textile spoof surface plasmon polariton (SSPP) waveguides offer guided
on-body links, but their body-facing ground conductors obstruct coupling
from on-skin wireless nodes. This letter presents a split-ring coupling
interface using local ground recesses at the textile terminations, without
patterned garment-side launchers. The capacitively loaded rings are supported
by polydimethylsiloxane carriers with a thickness of 2~mm. An image-current model and full-wave
simulations examine the effect of backing-ground spacing and determine the
recess geometry. For a 450~mm textile waveguide on a water-filled body
phantom, the measured $S_{21}$ at 2.45 GHz was $-$18.8~dB, versus
$-$60.9~dB without the textile. A 625-position scan showed a maximum
5.1 dB reduction relative to the nominal position within the measured joint
$\pm$5~mm placement region. A separate two-node IEEE~802.11n test demonstrated
a 65~Mb/s physical-layer (PHY) rate with zero packet errors among 300 frames at
$-$20~dBm transmit power through the coupler link. These results establish high-rate, placement-tolerant on-skin Wi-Fi communication through an SSPP waveguide textile.
\end{abstract}


\section{Introduction}\label{sec:intro}

\IEEEPARstart{S}{kin-side} wearable nodes can communicate through radiative
links, but nearby tissue may increase propagation loss and detune antennas
\cite{Hall_TAP_2007}. Textile spoof surface plasmon polariton (SSPP)
waveguides guide microwave signals along periodically corrugated conductors
\cite{Pendry_Science_2004,Shen2013Conformal,Tian2020Conformal,Maier_PRL_2006}; a review is given in \cite{GarciaVidal_RMP_2022}.
A body-facing ground conductor reduces interaction with tissue
\cite{Zhu2024Robust}. Prior textile-based systems have addressed sensing and low-power nodes
\cite{Tian2019Wireless,Lin2020Wireless,Kurt2026Battery}, multibody and
reconfigurable networks \cite{Hajiaghajani2021Textile,Tian2024Phased,Li2026Body},
and wireless powering and implant links
\cite{Agrawal2017Conformal,Tian2023Implant,Pham2025Wearable}.

On-skin nodes reside between the body and the garment
\cite{Kim2011Epidermal}. In a grounded textile SSPP structure, the ground
conductor separates these nodes from the guiding strip. Conventional
transitions typically excite the line from an accessible end or the outer
garment surface \cite{Ma_LPR_2014,Liao_JPD_2014,Kianinejad2016LowLoss,ElBacha2024Design}.
A compact interface is therefore needed to couple on-skin nodes without
adding a dedicated patterned transition to each textile endpoint.
Unlike narrowband sensor telemetry, Wi-Fi also requires a transmission
response that accommodates its modulated signal bandwidth.

This letter develops a ground-recessed termination and a capacitively loaded
split-ring coupler on a flexible carrier. The interface is characterized by
image-current analysis, full-wave simulation, and $S$-parameter measurements
on a phantom-loaded textile. Two-dimensional placement measurements quantify
sensitivity to node position, and an IEEE 802.11n link test verifies Wi-Fi signal
transmission through the proposed interface.

\section{Coupling Interface Design}\label{sec:structure}
\subsection{Interface Configuration and Coupling Principle}

\begin{figure}[!t]
\centering
\includegraphics[width=0.85\columnwidth]{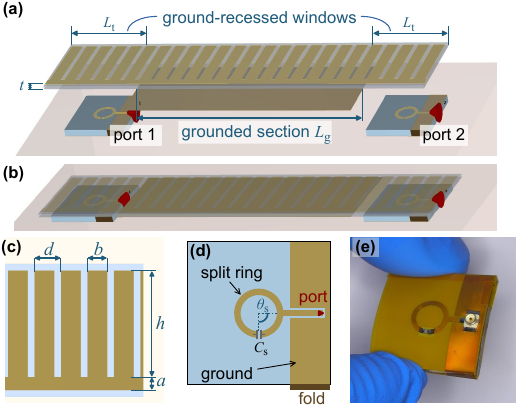}
\caption{(a) Exploded view of the simulation model, with a ground-recessed
window of length $L_\mathrm{t}$ at each termination and the grounded section
$L_\mathrm{g}$ between them. (b) The same model assembled. (c) Periodic comb structure, defining $a$, $b$, $d$ and $h$. (d) The coupler's waveguide-facing conductor
pattern; the fold carries the coplanar feed ground to the body-facing
face. (e) Fabricated coupler.}
\label{fig:geometry}
\end{figure}

\reffig{geometry} illustrates the interface, and \reftab{param}
summarizes its dimensions. The textile incorporates a periodic comb conductor
on its outer surface and a ground conductor on the body-facing surface. The ground is recessed by
$L_\mathrm{t}$ at each termination to expose the comb to a on-skin coupler.
The coupler comprises a split ring of mean radius $r_0$ on the
garment-facing surface of a \SI{2}{\milli\metre} polydimethylsiloxane (PDMS)
carrier, a loading
capacitor $C_\mathrm{s}$, a coplanar feed, and a backing ground on the
skin-facing surface. A folded conductor connects the backing plane to the
coplanar feed ground (\reffig{geometry}(d)).

\begin{table}[!t]
\renewcommand{\arraystretch}{1.12}
\caption{Design Values}
\label{tab:param}
\centering
\footnotesize
\begin{tabular}{@{}p{0.40\columnwidth}p{0.55\columnwidth}@{}}
\hline
SSPP unit cell $a/b/d/h/t$ & 2.5 / 3.75 / 5 / 20 / 1 mm\\
Waveguide width $w$ & \SI{25}{\milli\metre}\\
Waveguide length $L$ & \SI{125}{\milli\metre} (reference sim.); \SI{450}{\milli\metre} (prototype)\\
Ground recess $L_\mathrm{t}$ & \SI{25}{\milli\metre} (prototype); \SIrange{21}{22.5}{\milli\metre} (sim. optima)\\
Split ring $r_0/w_\mathrm{r}/\theta_\mathrm{s}$ & \SI{3.0}{\milli\metre} / \SI{0.9}{\milli\metre} / $90^\circ$\\
Loading capacitor $C_\mathrm{s}$ & \SI{0.30}{\pico\farad} (prototype); \SI{0.36}{\pico\farad} (sim. optimum)\\
PDMS spacer (size, $\varepsilon_\mathrm{r}$) & $20\times20\times\SI{2}{\milli\metre}$; 2.7\\
Coupler-to-waveguide gap $g$ & \SIrange{0.5}{3.0}{\milli\metre} (sim.); \SI{1}{\milli\metre} (nominal)\\
Simulation phantom & $\eps_r=40$, $\sigma=\SI{1.8}{\siemens\per\metre}$\\
\hline
\end{tabular}
\end{table}

\begin{figure}[!t]
\centering
\includegraphics[width=0.99\columnwidth]{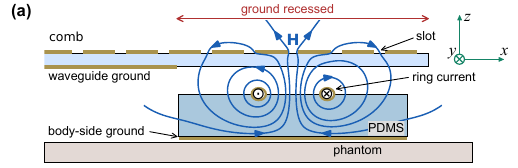}\\[2pt]
\input{figs/fig_hfield}
\caption{(a) Schematic of magnetic coupling at a recessed termination, sectioned through the ring diameter; $\odot$ and $\otimes$ indicate current flowing out of and into the page, respectively.
(b) Simulated $|\mathbf{H}|$ on the $y=0$ plane at
\SI{2.464}{\giga\hertz}, to full scale at \SI{62.8}{\ampere\per\metre},
the 99th percentile of the window; vertical scale $\times 2$. Dotted,
the coupler centers; dashed, the edges of the ground recession.}
\label{fig:mechanism}
\end{figure}

The ring, feed, and ground interconnect are patterned on a flexible
circuit folded around the PDMS carrier, avoiding a through-PDMS via. The simulated field in \reffig{mechanism}(b) is concentrated near the split ring and exposed comb, consistent with near-field coupling at the recessed termination.

At \SI{2.45}{\giga\hertz}, the ring circumference is approximately
$0.15\lambda_0$. The capacitor $C_\mathrm{s}$ controls resonance, whereas the
angular separation $\theta_\mathrm{s}$ between split and feed also influences
the input impedance. Simulated ring radii of
\SIrange{2.6}{3.4}{\milli\metre}, retuned individually, yielded a maximum
difference of \SI{0.09}{\decibel} in peak delivered power. Here, we selected $r_0=\SI{3.0}{\milli\metre}$.

\subsection{Coupler Design and Analysis}\label{sec:coupler}

Full-wave electromagnetic simulations were performed using CST Studio Suite.
The coupler's skin-facing backing ground is intended to reduce electromagnetic
coupling into the body, as does the ground of the textile SSPP waveguide.
In the complete simulation, removing the coupler backing ground
reduced $S_{21}$ by \SI{7.4}{\decibel} under the evaluated tuning conditions
and increased the fraction of accepted power absorbed by the phantom from
69\% to 87\%. These results support retaining the backing ground, although phantom
absorption remains non-negligible.

\begin{figure}[!t]
\centering
\input{figs/fig_imagecurrent}
\caption{(a) Simulated peak transmission and maximum accepted power fraction versus ring-to-ground separation $t_\mathrm{s}$; (b) Magnetic field per unit ring current on the ring axis at two heights, normalized to $t_\mathrm{s}=\SI{2}{\milli\metre}$. A \SI{1}{\ampere} source drives the split. Lines show \refeq{image} with $r_0=\SI{3.0}{\milli\metre}$, without fitting.}
\label{fig:image}
\end{figure}
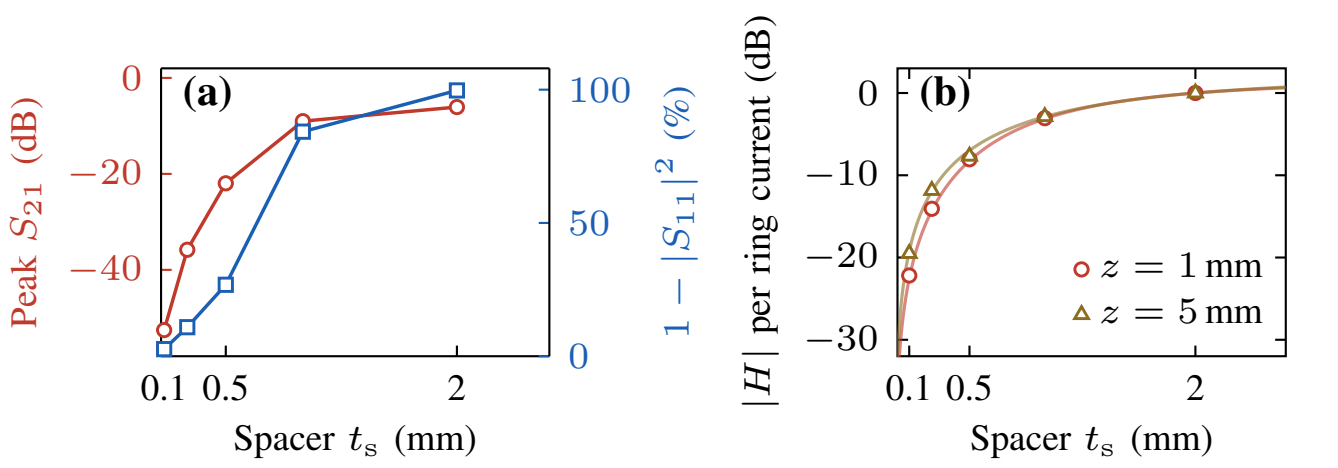

The ground must nevertheless be sufficiently separated from the ring to avoid
suppressing its magnetic field toward the waveguide. A conventional two-layer
flexible printed circuit (FPC) can provide a ring-to-ground separation of only
\SIrange{25}{50}{\micro\metre}. For a circular current loop parallel to a
perfect electric conductor, the method of images replaces the ground with an
oppositely directed image loop. Subtracting the axial fields of the real and
image loops, each proportional to $[r_0^2+z^2]^{-3/2}$ for fixed loop current
and radius, gives
\begin{equation}
H(z)\;\propto\;\bigl(r_0^{2}+z^{2}\bigr)^{-3/2}
-\bigl(r_0^{2}+(z+2t_\mathrm{s})^{2}\bigr)^{-3/2},
\label{eq:image}
\end{equation}
where $r_0$ is the loop radius, $t_\mathrm{s}$ is the ring-to-backing-ground
separation, and $z$ is the observation distance from the ring plane toward
the textile waveguide. The image loop is located $2t_\mathrm{s}$ behind the
ring plane.

Image-current cancellation was evaluated in a simplified model containing
only the ring, its dielectric carrier, and the backing plane; the loading
capacitor, the coplanar feed ground, the edge fold, and the feed trace were
removed. An ideal \SI{1}{\ampere} source at the split fixed the ring current. Without fitted parameters, the analytical model agreed with simulations
within \SI{1.27}{\decibel} over a \SI{30.8}{\decibel} field range,
(\reffig{image}(b)). Equation~\eqref{eq:image} predicts a
\SI{28.7}{\decibel} reduction in magnetic field per unit ring current for
$t_\mathrm{s}=\SI{50}{\micro\metre}$ relative to $t_\mathrm{s}=\SI{2}{\milli\metre}$. Thus, the nearby backing ground substantially cancels the ring field at small separations.

The complete two-coupler simulation shows the same thickness dependence
(\reffig{image}(a)): decreasing $t_\mathrm{s}$ suppresses transmission
and reduces the power accepted by the coupler. The agreement between the
image-current model and simulated fields in \reffig{image}(b) identifies
field cancellation by the nearby backing ground as an explanation
for the observed loss of coupling with thinner spacers. We therefore used a \SI{2}{\milli\metre} PDMS
spacer for the present geometry; the result does not exclude other thin-substrate couplers.

Standing-wave effects in the complete two-coupler model varied $S_{21}$ at
\SI{2.45}{\giga\hertz} by \SI{4.0}{\decibel} across the simulated guide
lengths. To separate coupling from reflections at the second coupler during coupler design optimization, we used
a single-coupler model with graded absorbing terminations, omitting the
phantom, waveguide backing ground, and second coupler. The fraction $\kappa$
of accepted power reaching the absorbers was stable (\SIrange{87.1}{87.9}{\percent})
over guide lengths of \SIrange{145}{185}{\milli\metre}, which span more than
one guided wavelength, supporting comparative optimization of the coupler
geometry.

In the single coupler simulation, both $C_\mathrm{s}$ and $\theta_\mathrm{s}$ affect resonance and input
matching. The simulated $|S_{11}|$ minimum varied by more than
\SI{400}{\mega\hertz} over the evaluated capacitor range and by up to
\SI{95}{\mega\hertz} with split position. Parameter combinations yielding
resonance near \SI{2.45}{\giga\hertz} were compared across coupling gaps of
\SIrange{0.5}{3.0}{\milli\metre}. The simulated optimum was
$C_\mathrm{s}=\SI{0.36}{\pico\farad}$ and $\theta_\mathrm{s}=90^\circ$ under
the adopted worst-gap objective.

A nominal \SI{0.30}{\pico\farad} capacitor was installed in the prototype.
The chosen discrete value differs from the simulated optimum; mounted
parasitics were not independently measured.

\section{Ground-Recess Design}\label{sec:recess}

The ground-recess length $L_\mathrm{t}$ affects transmission and
longitudinal placement sensitivity.

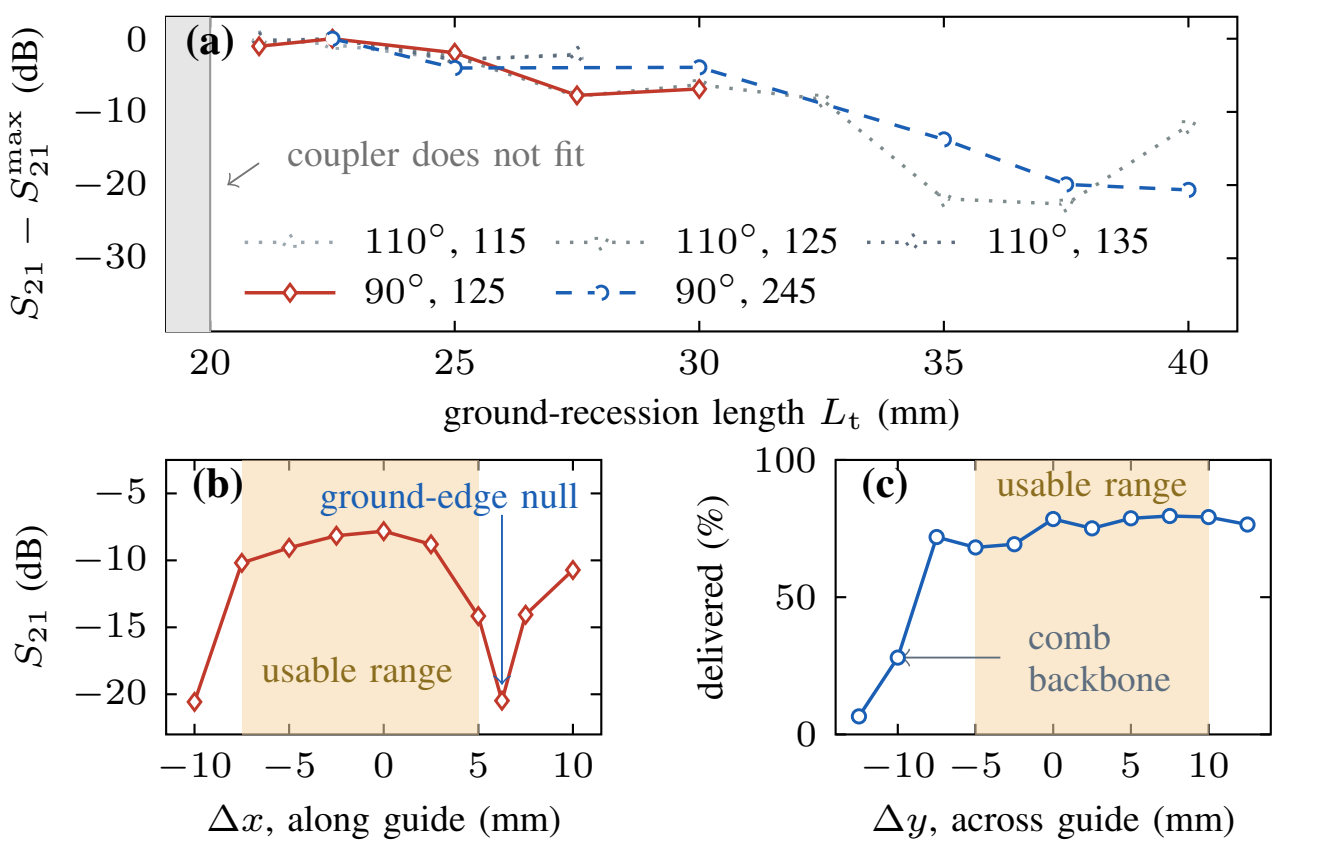
\begin{figure}[!t]
\centering
\input{figs/fig_tolerance}
\caption{Simulation. (a) Transmission versus ground-recession length for five
cases, labeled by split offset $\theta_\mathrm{s}$ and guide length $L$ in mm;
each curve is referred to its own maximum, so \SI{0}{\decibel} marks that
case's optimum. Gray, recesses shorter than the \SI{20}{\milli\metre} coupler
footprint. (b) Longitudinal placement, complete model at
$L_\mathrm{t}=\SI{22.5}{\milli\metre}$. (c) Transverse placement, terminated
model, in which \emph{delivered} is the share of incident power that reaches
the guide. Gold in (b) and (c), the usable placement range cited in the text;
the two models are not comparable in magnitude.}
\label{fig:tol}
\end{figure}
\reffig{tol}(a) compares five simulated configurations with different guide
lengths and coupler arrangements. A \SI{22.5}{\milli\metre} recess maximized
transmission in most cases. Increasing $L_\mathrm{t}$ generally introduced additional loss. An exception near
\SI{40}{\milli\metre} occurred for the shortest guide, consistent with a
length-dependent standing-wave response. These results favor shorter recesses within the investigated configurations;
they do not establish an optimum for other guide lengths.

The longitudinal sweep in \reffig{tol}(b) exhibits a transmission minimum
approximately \SI{6.25}{\milli\metre} inward from the recess center, with a
depth of about \SI{13}{\decibel}. Its location follows approximately
$L_\mathrm{t}/2-\SI{5}{\milli\metre}$ over the cases examined. Shifting the
nominal coupler position outward by \SI{2.25}{\milli\metre} increased the
minimum transmission within a $\pm\SI{5}{\milli\metre}$ longitudinal interval
by \SI{5.1}{\decibel}.

The transverse displacement response in \reffig{tol}(c) is asymmetric because
displacement toward the comb backbone and toward the teeth produces different
overlaps. The evaluated usable range was approximately $-5$ to
$+\SI{10}{\milli\metre}$ in the transverse direction, compared with $-7.5$ to
$+\SI{5}{\milli\metre}$ longitudinally at
$L_\mathrm{t}=\SI{22.5}{\milli\metre}$. The sweeps use different simulation models and cannot establish a joint
two-dimensional tolerance. No consistent tolerance improvement was observed
for longer recesses in the investigated cases.

The fabricated recess length was constrained to integer multiples of the \SI{5}{\milli\metre} comb period. A \SI{20}{\milli\metre} recess provides insufficient clearance for the \SI{20}{\milli\metre} coupler footprint. We therefore selected five periods (\SI{25}{\milli\metre}) as the shortest feasible recess. Relative to the continuously optimized recess, this choice incurred simulated transmission penalties of 1.9 and \SI{4.0}{\decibel} for guide lengths of 125 and \SI{245}{\milli\metre}, respectively.

\section{Experimental Validation}\label{sec:meas}

Two couplers were fabricated with the dimensions in \reftab{param}
(\reffig{geometry}(e)). The fabricated SSPP waveguide was \SI{450}{\milli\metre} long,
with a \SI{25}{\milli\metre} recess and a nominal separation of
\SI{425}{\milli\metre} between couplers. The textile was placed over the
couplers on the exterior of a water-filled container (\reffig{meas}(a)).
The measurement setup and placement scan are documented in the supplementary video.

\begin{figure}[!t]
\centering
\input{figs/fig_measurement}
\vspace{-5mm}  %
\caption{(a) Measurement schematic: the textile SSPP waveguide wrapped around a
water-filled container, a coupler at each end of the comb, each cabled to a
vector network analyzer (VNA) port. (b) The same, photographed; the couplers are \SI{425}{\milli\metre}
apart. (c) Measured $|S_{21}|$ of the link and of the same coupler pair with
the textile removed; the shaded band is \SIrange{2.400}{2.4835}{\giga\hertz}.
(d) Measured $|S_{21}|$ against coupler displacement, best of wireless local area network (WLAN) channels 1,
6 and 11; $\Delta x$ along the waveguide, $\Delta y$ across it; contours 3 and
\SI{6}{\decibel} below the center value of $\mapCenter$~dB; dashed square,
the $\pm\SI{5}{\milli\metre}$ envelope. (e) Worst case of the same quantity
over each envelope, on the scale of (d).}
\label{fig:meas}
\end{figure}
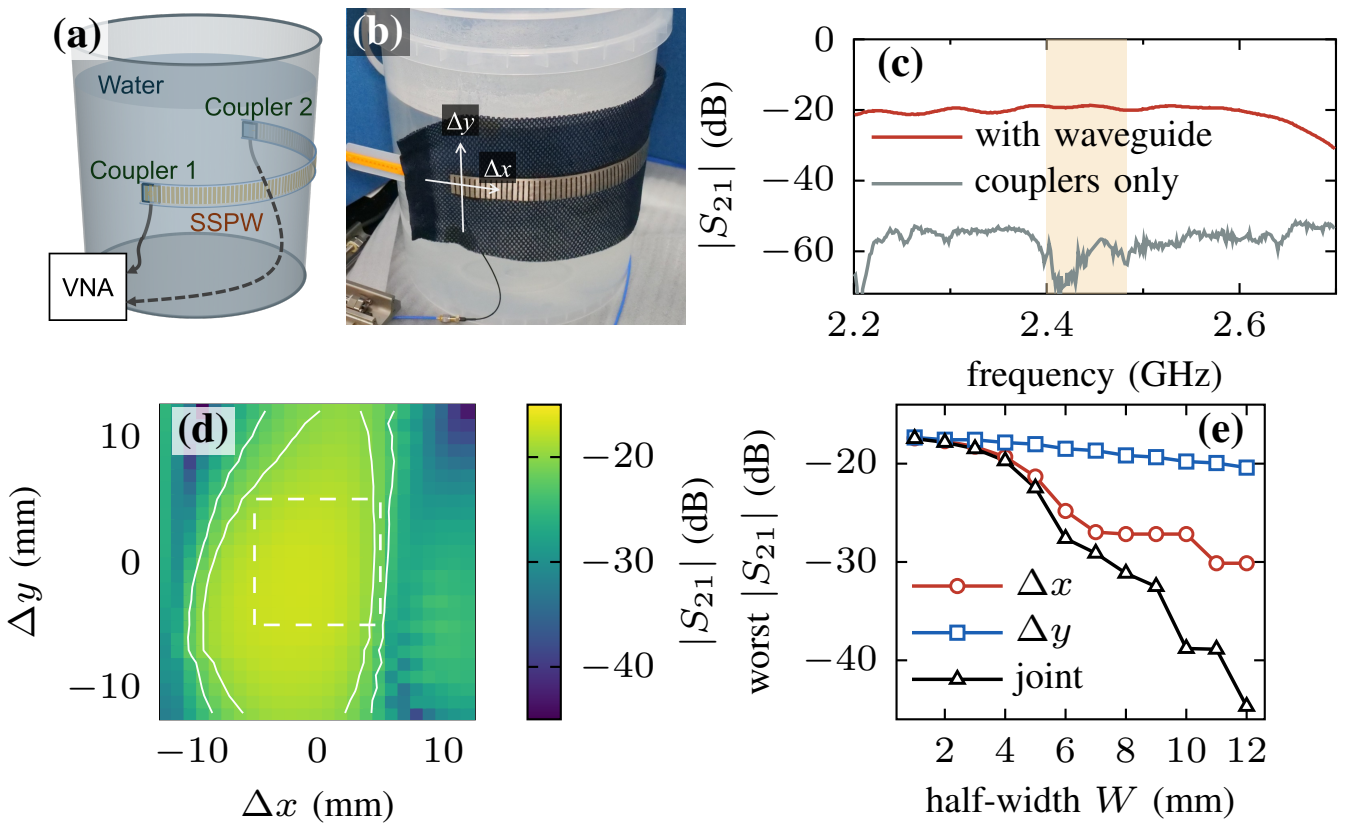

A two-port vector network analyzer (VNA) measured the response after
unknown-through/open/short/match (UOSM) calibration,
using a $-10$-dBm source level and a \SI{10}{\kilo\hertz} intermediate-frequency (IF) bandwidth.

At \SI{2.45}{\giga\hertz}, the measured $S_{21}$ magnitude was $-18.8$~dB with
the textile waveguide and $-60.9$~dB after removing the textile, with the coupler and cable
positions otherwise unchanged (\reffig{meas}(c)). The textile waveguide thus increased the measured $S_{21}$ by
\SI{42.1}{\decibel} relative to the no-textile control. The measured $S_{21}$ remained within \SI{3}{\decibel} of its maximum
from 2.20 to \SI{2.63}{\giga\hertz}. 

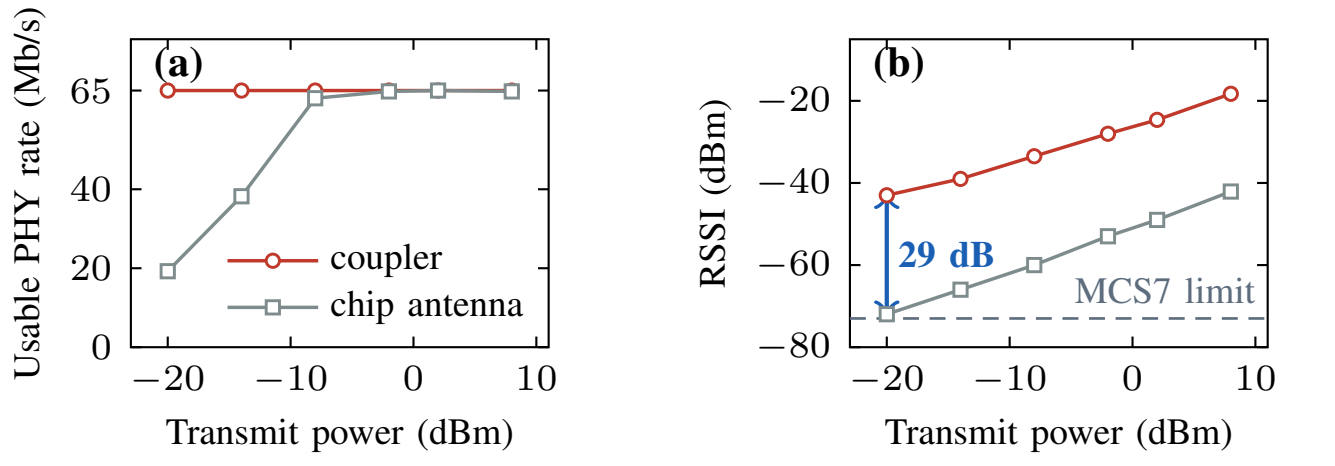
\begin{figure}[!t]
\centering
\input{figs/fig_wlan}
\caption{XIAO ESP32C6 boards using onboard antennas (INT) or U.FL-connected textile couplers (EXT): (a) highest physical-layer (PHY) rate with packet error rate (PER) below 10\%; (b) received signal strength. Tests used 20-MHz 802.11n on channel 9 with 1000-byte frames.}
\label{fig:wifi}
\end{figure}
Placement tolerance was measured on the same assembly. One coupler was
translated over $\pm\SI{12}{\milli\metre}$ on two axes at
\SI{1}{\milli\metre} pitch, giving 625 positions, with $\Delta x$ along the
waveguide and $\Delta y$ across it. 
\reffig{meas}(d) maps the best of WLAN
channels 1, 6 and 11 at each position, $\mapCenter$~dB at the center.
The wall is curved along $\Delta x$ and inclined along $\Delta y$, so contact
was exact only at that position despite a compliant support, and the scan
combines displacement with an uncontrolled variation of the coupling gap.

Transmission varied by 4.1 and \SI{9.9}{\decibel} over longitudinal intervals
of $\pm5$ and $\pm\SI{10}{\milli\metre}$, against 0.8 and
\SI{2.6}{\decibel} transversely. Two minima flanked the central plateau,
\SI{19.0}{\milli\metre} apart and 12.9 and \SI{9.9}{\decibel} below the
center, that is \SI{3.0}{\milli\metre} inside the displacements at which the
coupler center reaches the recess edge and the textile end; the corresponding
minimum in \reffig{tol}(b) lies \SI{5}{\milli\metre} inside the recess edge.
The worst case over a joint $\pm\SI{5}{\milli\metre}$ square was $\mapWorst$~dB
(\reffig{meas}(e)), \SI{1.2}{\decibel} below the worse of the two cuts; at
$\pm\SI{10}{\milli\metre}$ the difference grew to \SI{11.6}{\decibel}.
Thus, one-dimensional sweeps understate the joint placement penalty. 
Reducing the same scan
as a mean over \SIrange{2.4}{2.5}{\giga\hertz}, rather than as the best of
the three channels, lowers the center value by \mapAltCenter~dB and the
$\pm\SI{5}{\milli\metre}$ worst case by \mapAltWorst~dB.
For reproducibility, a repeated scan after rewrapping retained the \SI{19.0}{\milli\metre}
separation, with a \SI{3.0}{\milli\metre} pattern shift and a
\SI{1.8}{\decibel} lower center value.

The inward-side transmission drop lies closer to the nominal center
than the outward-side drop (\reffig{meas}(d)). This trend is consistent
with the simulation in \reffig{tol}(b), where shifting the nominal
position outward improves minimum transmission over a
$\pm\SI{5}{\milli\metre}$ interval. Although the measured
$S_{21}$ at \SI{2.45}{\giga\hertz} is \SI{5.2}{\decibel} below
simulation, the broad, relatively flat frequency response and the
longitudinal placement trend agree qualitatively with the simulated
behavior. Conductor loss and deviations from the optimized recess
and capacitance may contribute to the absolute-level difference.

Two Seeed Studio XIAO ESP32C6 boards \cite{SeeedXIAOESP32C6}
were used for 802.11n transmission tests (\reffig{wifi}). Their
software-controlled RF switches selected the onboard ceramic antennas
(INT) or the couplers connected through U.FL ports (EXT). Each test used
300 1000-byte frames on 20-MHz channel 9 (2.452 GHz). At $-20$ dBm,
MCS7 (65 Mb/s PHY rate) yielded zero packet errors through EXT, whereas
INT could not establish an MCS7 link. The rate is PHY, not application
throughput; a multi-node network was not evaluated.

\section{Conclusion}\label{sec:conclusion}

On-skin split-ring couplers enable Wi-Fi communication via a ground-recessed textile SSPP waveguide without garment-side launchers. The phantom-loaded prototype improved $S_{21}$ by 42.1 dB over the no-textile control and showed no deep transmission nulls within the sampled joint $\pm$5-mm placement region. A two-node 802.11n test confirmed packet transmission at a 65-Mb/s PHY rate using practical Wi-Fi transceivers.

\newpage
\bibliographystyle{IEEEtran}
\bibliography{references}

\end{document}

%% file: Macro.tex
\usepackage{amsmath,amssymb,amsfonts}
\usepackage{array}
\usepackage{cite}
\usepackage{graphicx}
\usepackage{flafter}
\usepackage[caption=false]{subfig}
\usepackage{multirow}
\usepackage{xcolor}
\usepackage{siunitx}
\def\eps{\varepsilon}
\newcommand{\reffig}[1]{Fig.~\ref{fig:#1}}
\newcommand{\reftab}[1]{Table~\ref{tab:#1}}
\newcommand{\refeq}[1]{(\ref{eq:#1})}

\newif\ifTBDvisible
\TBDvisibletrue

%% file: figs/plotstyle.tex
\usepackage{pgfplots}
\pgfplotsset{compat=1.17}
\usepgfplotslibrary{colormaps}
\usetikzlibrary{calc}

\definecolor{cRed}{HTML}{C0392B}
\definecolor{cBlue}{HTML}{1A5FB4}
\definecolor{cGold}{HTML}{8A6D1F}
\definecolor{cGrey}{HTML}{7F8C8D}
\definecolor{cGreyL}{HTML}{9AA7B1}
\definecolor{cGreyD}{HTML}{5D6D7E}
\definecolor{cBand}{HTML}{F6D7A7}
\definecolor{cFlexC}{HTML}{CFD8E3}

\pgfplotsset{
  paper/.style={
    scale only axis,
    /tikz/font=\scriptsize,
    label style={font=\scriptsize, inner sep=1pt},
    tick label style={font=\scriptsize},
    legend style={font=\scriptsize, draw=none, fill=none, inner sep=1pt,
                  row sep=-1.5pt, legend cell align=left},
    axis line style={line width=0.4pt},
    tick style={line width=0.4pt, color=black},
    major tick length=2pt,
    minor tick length=1pt,
    every axis plot/.append style={line width=0.65pt},
  },
  series/.style={mark size=1.25pt, mark options={fill=white, line width=0.55pt}},
  model/.style={no marks, line width=0.6pt, opacity=0.6},
  band/.style={draw=none, fill=cBand, fill opacity=0.55},
  layer/.style={white, line width=0.3pt, opacity=0.85, no marks},
}

\newcommand{\panel}[2][0.012]{%
  \node[anchor=north west, inner sep=1pt, font=\footnotesize\bfseries]
    at (rel axis cs:#1,0.995) {(#2)};}

\newcommand{\panelw}[2][0.012]{%
  \node[anchor=north west, inner sep=1pt, font=\footnotesize\bfseries,
        fill=white, fill opacity=0.75, text opacity=1]
    at (rel axis cs:#1,0.995) {(#2)};}

\input{figs/field_extent}

\input{figs/wlan_values}
\input{figs/meas_values}

\newcommand{\plotdata}{data/plot}

%% file: figs/field_extent.tex
\newcommand{\hfXmin}{-72.75}
\newcommand{\hfXmax}{72.75}
\newcommand{\hfZmin}{-2.25}
\newcommand{\hfZmax}{6.25}

\newcommand{\hfQuant}{$|\mathbf{H}|$}
\newcommand{\hfFile}{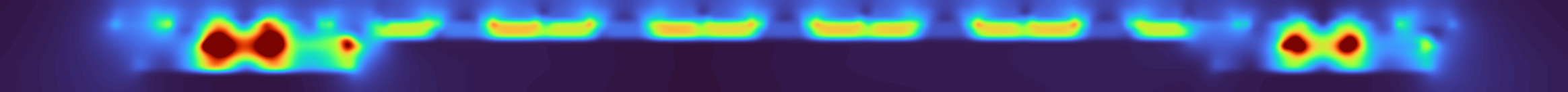}
\newcommand{\hfCbar}{./Figures/field_cbar.png}

%% file: figs/wlan_values.tex
\newcommand{\wlGapX}{-20.00}
\newcommand{\wlGapLo}{-71.97}
\newcommand{\wlGapHi}{-43.00}
\newcommand{\wlGap}{29}
\newcommand{\wlSens}{-73.0}

%% file: figs/meas_values.tex
\newcommand{\mapCenter}{-17.3}

\newcommand{\mapWorst}{-22.5}
\newcommand{\mapAltCenter}{0.5}
\newcommand{\mapAltWorst}{1.0}

%% file: figs/fig_hfield.tex
\begin{tikzpicture}
\pgfplotsset{paper}
\begin{axis}[name=hf, width=0.855\columnwidth, height=0.0999\columnwidth,
             xmin=\hfXmin, xmax=\hfXmax, ymin=\hfZmin, ymax=\hfZmax,
             xtick={-60,-40,-20,0,20,40,60}, ytick={0,2,4,6},
             xlabel={$x$ (mm)}, ylabel={$z$ (mm)},
             x label style={inner sep=0pt}, y label style={inner sep=0pt},
             axis on top, tick align=outside]
  \addplot graphics [xmin=\hfXmin, xmax=\hfXmax,
                     ymin=\hfZmin, ymax=\hfZmax] {\hfFile};
  \addplot[layer] coordinates {(-62.5,4) (62.5,4)};
  \addplot[layer] coordinates {(-37.5,3) (37.5,3)};
  \foreach \s in {-1,1} \foreach \zz in {0,2}
    \addplot[layer] coordinates {(\s*40,\zz) (\s*60,\zz)};
  \foreach \xx in {-37.5,37.5}
    \addplot[white, line width=0.4pt, densely dashed, no marks]
      coordinates {(\xx,\hfZmin) (\xx,\hfZmax)};
  \foreach \xx in {-50,50}
    \addplot[white, line width=0.4pt, densely dotted, no marks]
      coordinates {(\xx,\hfZmin) (\xx,\hfZmax)};
\end{axis}
\node[anchor=south east, inner sep=0pt] (cb)
  at ($(hf.north east) + (-0.024\columnwidth,0.022\columnwidth)$)
  {\includegraphics[width=0.24\columnwidth, height=0.018\columnwidth]{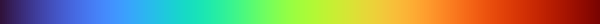}};
\draw[line width=0.3pt] (cb.south west) rectangle (cb.north east);
\node[anchor=east, font=\scriptsize, inner sep=1.6pt] at (cb.west) {0};
\node[anchor=west, font=\scriptsize, inner sep=1.6pt] at (cb.east) {1};
\node[anchor=east, font=\scriptsize, inner sep=1.6pt]
  at ($(cb.west) - (0.036\columnwidth,0)$) {\hfQuant{} (a.u.)};
\node[anchor=north west, inner sep=1pt, font=\footnotesize\bfseries]
  at ($(hf.north west) + (-0.075\columnwidth,0.045\columnwidth)$) {(b)};
\end{tikzpicture}

%% file: figs/fig_imagecurrent.tex
\begin{tikzpicture}
\pgfplotsset{paper, width=0.29\columnwidth, height=0.215\columnwidth,
             xmode=log, xtick={0.1,0.5,2}, xticklabels={0.1,0.5,2},
             log ticks with fixed point, xlabel={Spacer $t_\mathrm{s}$ (mm)}}

\begin{axis}[name=la, xmin=0.08, xmax=2.6, ymin=-58, ymax=2,
             ytick={0,-20,-40}, ylabel={Peak $S_{21}$ (dB)},
             y label style={text=cRed}, ytick style={color=cRed},
             yticklabel style={text=cRed}, axis y line*=left,
             axis x line*=bottom]
  \addplot[cRed, series, mark=*] table[x=h, y=peak]
    {\plotdata/gnd_spacing.dat};
  \panel[0.04]{a}
\end{axis}
\begin{axis}[name=lb, at={(la.south west)}, anchor=south west,
             xmin=0.08, xmax=2.6, ymin=0, ymax=108, ytick={0,50,100},
             ylabel={$1-|S_{11}|^2$ (\%)}, y label style={text=cBlue},
             ytick style={color=cBlue}, yticklabel style={text=cBlue},
             axis y line*=right, axis x line*=top, xtick=\empty,
             xminorticks=false, xlabel={}]
  \addplot[cBlue, series, mark=square*] table[x=h, y=accepted]
    {\plotdata/gnd_spacing.dat};
\end{axis}

\begin{axis}[name=rb, at={(la.south east)}, xshift=0.26\columnwidth,
             anchor=south west, xmin=0.022, xmax=2.6, ymin=-32, ymax=3,
             ytick={0,-10,-20,-30},
             ylabel={$|H|$ per ring current (dB)},
             legend style={at={(0.985,0.03)}, anchor=south east}]
  \addplot[cRed, model, forget plot]
    table[x=h, y=pred] {\plotdata/ring_image_model_z1.dat};
  \addplot[cGold, model, forget plot]
    table[x=h, y=pred] {\plotdata/ring_image_model_z5.dat};
  \addplot[cRed, series, only marks, mark=*]
    table[x=h, y=z1] {\plotdata/ring_image_meas.dat};
  \addlegendentry{$z=1$\,mm}
  \addplot[cGold, series, only marks, mark=triangle*, mark size=1.5pt]
    table[x=h, y=z5] {\plotdata/ring_image_meas.dat};
  \addlegendentry{$z=5$\,mm}
  \panel[0.04]{b}
\end{axis}
\end{tikzpicture}

%% file: figs/fig_tolerance.tex
\begin{tikzpicture}
\pgfplotsset{paper}

\begin{axis}[name=top, width=0.80\columnwidth, height=0.235\columnwidth,
             xmin=19.1, xmax=41, ymin=-40, ymax=3,
             xtick={20,25,30,35,40}, ytick={0,-10,-20,-30},
             xlabel={ground-recession length $L_\mathrm{t}$ (mm)},
             ylabel={$S_{21}-S_{21}^{\max}$ (dB)},
             legend style={at={(0.5,0.01)}, anchor=south, legend columns=3,
                           font=\scriptsize, /tikz/column sep=3pt}]
  \fill[black!10] (axis cs:19.1,-40) rectangle (axis cs:20,3);
  \draw[black!40, line width=0.4pt] (axis cs:20,-40) -- (axis cs:20,3);
  \draw[->, black!55, line width=0.35pt]
    (axis cs:21.0,-17.0) -- (axis cs:20.35,-20.0);
  \node[black!55, font=\scriptsize, anchor=west]
    at (axis cs:21.2,-16.2) {coupler does not fit};
  \addplot[cGreyL, series, mark=triangle*, mark size=1.5pt, dotted]
    table[x=gl, y=rel] {\plotdata/gnd_recession_oldL115.dat};
  \addlegendentry{$110^\circ$, 115}
  \addplot[cGrey, series, mark=triangle*, mark size=1.5pt,
           mark options={rotate=180, fill=white, line width=0.55pt}, dotted]
    table[x=gl, y=rel] {\plotdata/gnd_recession_oldL125.dat};
  \addlegendentry{$110^\circ$, 125}
  \addplot[cGreyD, series, mark=diamond*, mark size=1.6pt, dotted]
    table[x=gl, y=rel] {\plotdata/gnd_recession_oldL135.dat};
  \addlegendentry{$110^\circ$, 135}
  \addplot[cRed, series, mark=diamond*, mark size=1.6pt]
    table[x=gl, y=rel] {\plotdata/gnd_recession_newL125.dat};
  \addlegendentry{$90^\circ$, 125}
  \addplot[cBlue, series, mark=*, dashed]
    table[x=gl, y=rel] {\plotdata/gnd_recession_newL245.dat};
  \addlegendentry{$90^\circ$, 245}
  \panel{a}
\end{axis}

\begin{axis}[name=bl, at={(top.south west)}, anchor=north west,
             yshift=-8.5mm, width=0.325\columnwidth, height=0.205\columnwidth,
             xmin=-11.5, xmax=11.5, ymin=-23, ymax=-2.5,
             xtick={-10,-5,0,5,10}, ytick={-20,-15,-10,-5},
             xlabel={$\Delta x$, along guide (mm)}, ylabel={$S_{21}$ (dB)}]
  \fill[cBand, fill opacity=0.55] (axis cs:-7.5,-23) rectangle
    (axis cs:5,-2.5);
  \node[cGold, font=\scriptsize, anchor=center] at (axis cs:-1.4,-18.4)
    {usable range};
  \addplot[cRed, series, mark=diamond*, mark size=1.6pt]
    table[x=dx, y=s21] {\plotdata/placement_x.dat};
  \draw[->, cBlue, line width=0.35pt]
    (axis cs:6.25,-6.6) -- (axis cs:6.25,-19.3);
  \node[cBlue, font=\scriptsize, anchor=north east]
    at (axis cs:11.3,-2.8) {ground-edge null};
  \panel[0.045]{b}
\end{axis}

\begin{axis}[name=br, at={(bl.south east)}, anchor=south west,
             xshift=0.175\columnwidth,
             width=0.325\columnwidth, height=0.205\columnwidth,
             xmin=-14, xmax=14, ymin=0, ymax=100,
             xtick={-10,-5,0,5,10}, ytick={0,50,100},
             xlabel={$\Delta y$, across guide (mm)},
             ylabel={delivered (\%)}]
  \fill[cBand, fill opacity=0.55] (axis cs:-5,0) rectangle (axis cs:10,100);
  \node[cGold, font=\scriptsize, anchor=center] at (axis cs:2.5,90)
    {usable range};
  \addplot[cBlue, series, mark=*]
    table[x=dy, y=delivered] {\plotdata/placement_y.dat};
  \draw[->, cGreyD, line width=0.35pt]
    (axis cs:-3.4,28) -- (axis cs:-9.8,28);
  \node[cGreyD, font=\scriptsize, align=left, anchor=west]
    at (axis cs:-2.8,28) {comb\\backbone};
  \panel[0.045]{c}
\end{axis}
\end{tikzpicture}

%% file: figs/fig_measurement.tex
\begin{tikzpicture}
\pgfplotsset{paper}

\begin{axis}[name=spec, width=0.36\columnwidth, height=0.19\columnwidth,
             xmin=2.2, xmax=2.7, ymin=-72, ymax=0,
             xtick={2.2,2.4,2.6}, minor xtick={2.3,2.5,2.7},
             ytick={0,-20,-40,-60},
             xlabel={frequency (GHz)}, ylabel={$|S_{21}|$ (dB)},
             legend style={at={(0.02,0.30)}, anchor=south west}]
  \fill[cBand, fill opacity=0.45] (axis cs:2.4,-72) rectangle
    (axis cs:2.4835,0);
  \addplot[cRed, no marks] table[x=f, y=s21] {\plotdata/s21_with.dat};
  \addlegendentry{with waveguide}
  \addplot[cGrey, no marks] table[x=f, y=s21] {\plotdata/s21_wo.dat};
  \addlegendentry{couplers only}
  \panel[0.035]{c}
\end{axis}

\node[anchor=east, inner sep=0pt] (photo)
  at ($(spec.west) + (-0.125\columnwidth,0)$)
  {\includegraphics[width=0.253\columnwidth]{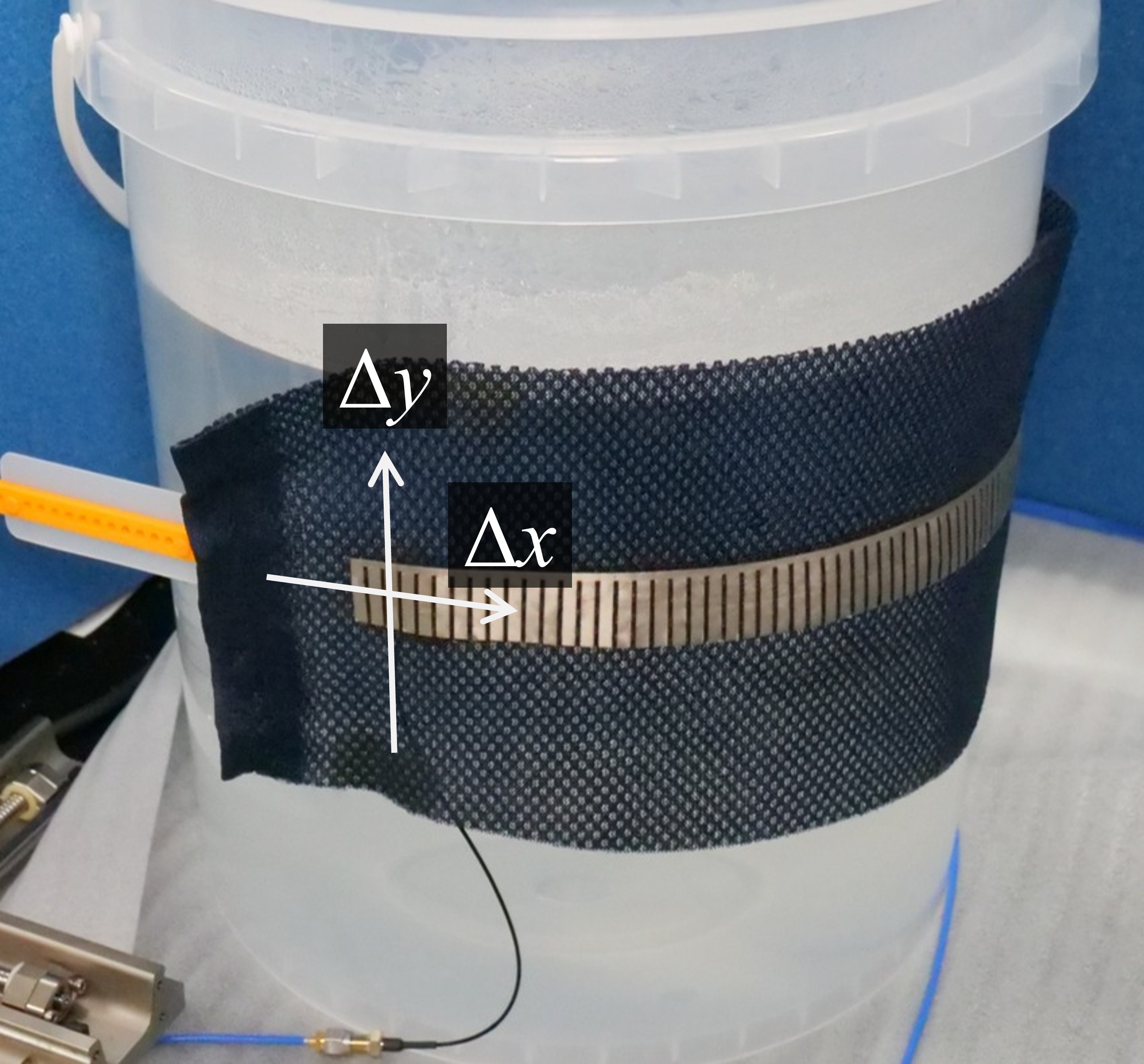}};
\node[anchor=north west, inner sep=1pt, font=\footnotesize\bfseries,
      text=white, fill=black, fill opacity=0.45, text opacity=1]
  at (photo.north west) {(b)};
\node[anchor=east, inner sep=0pt] (draw)
  at ($(photo.west) + (-0.012\columnwidth,0)$)
  {\includegraphics[width=0.210\columnwidth]{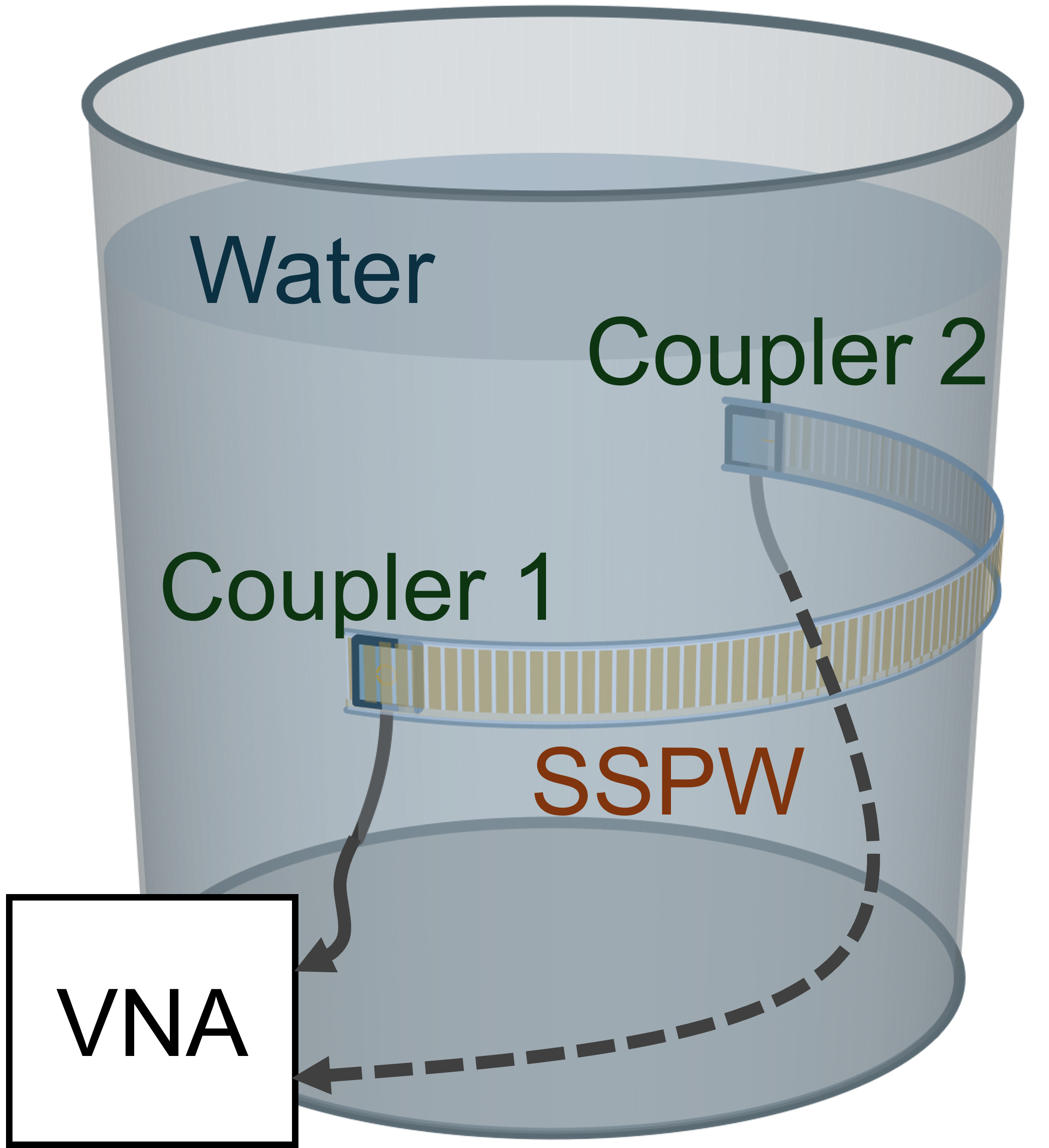}};
\node[anchor=north west, inner sep=1pt, font=\footnotesize\bfseries,
      fill=white, fill opacity=0.75, text opacity=1]
  at (draw.north west) {(a)};

\begin{axis}[name=map, at={(draw.south west)}, anchor=north west,
             xshift=0.085\columnwidth, yshift=-5.5mm,
             width=0.235\columnwidth, height=0.235\columnwidth,
             xmin=-12.5, xmax=12.5, ymin=-12.5, ymax=12.5,
             xtick={-10,0,10}, ytick={-10,0,10},
             xlabel={$\Delta x$ (mm)}, ylabel={$\Delta y$ (mm)},
             colormap/viridis, point meta min=-45, point meta max=-15,
             colorbar, colorbar style={width=2.2mm, ytick={-20,-30,-40},
                                       ylabel={$|S_{21}|$ (dB)},
                                       y label style={inner sep=1pt},
                                       /tikz/font=\scriptsize}]
  \addplot[matrix plot*, mesh/cols=25, point meta=explicit]
    table[x=x, y=z, meta=s21] {\plotdata/placement_map.dat};
  \foreach \lv in {3,6} {
    \foreach \k in {0,1} {
      \addplot[white, no marks, line width=0.35pt]
        table[x=x, y=z] {\plotdata/map_contour_\lv_\k.dat};
    }
  }
  \draw[white, dashed, line width=0.5pt]
    (axis cs:-5,-5) rectangle (axis cs:5,5);
  \panelw[0.035]{d}
\end{axis}

\begin{axis}[name=guar, at={(map.south east)}, anchor=south west,
             xshift=0.315\columnwidth,
             width=0.275\columnwidth, height=0.235\columnwidth,
             xmin=0.4, xmax=12.6, ymin=-46, ymax=-14,
             xtick={2,4,6,8,10,12}, ytick={-20,-30,-40},
             xlabel={half-width $W$ (mm)},
             ylabel={worst $|S_{21}|$ (dB)},
             legend style={at={(0.02,0.02)}, anchor=south west,
                           legend columns=1}]
  \addplot[cRed, series, mark=*] table[x=W, y=lon] {\plotdata/guarantee.dat};
  \addlegendentry{$\Delta x$}
  \addplot[cBlue, series, mark=square*]
    table[x=W, y=tra] {\plotdata/guarantee.dat};
  \addlegendentry{$\Delta y$}
  \addplot[black, series, mark=triangle*, mark size=1.5pt]
    table[x=W, y=joint] {\plotdata/guarantee.dat};
  \addlegendentry{joint}
  \panel[0.80]{e}
\end{axis}
\end{tikzpicture}

%% file: figs/fig_wlan.tex
\begin{tikzpicture}
\pgfplotsset{paper, width=0.312\columnwidth, height=0.23\columnwidth,
             xmin=-23, xmax=11, xtick={-20,-10,0,10},
             xlabel={Transmit power (dBm)}}

\begin{axis}[name=rate, ymin=0, ymax=78, ytick={0,20,40,65},
             ylabel={Usable PHY rate (Mb/s)},
             legend style={at={(0.98,0.02)}, anchor=south east}]
  \addplot[cRed, series, mark=*] table[x=tx, y=rate]
    {\plotdata/wlan_rate_cpl.dat};
  \addlegendentry{coupler}
  \addplot[cGrey, series, mark=square*] table[x=tx, y=rate]
    {\plotdata/wlan_rate_ant.dat};
  \addlegendentry{chip antenna}
  \panel[0.04]{a}
\end{axis}

\begin{axis}[name=rssi, at={(rate.south east)}, anchor=south west,
             xshift=0.225\columnwidth, ymin=-80, ymax=-5,
             ytick={-80,-60,-40,-20}, ylabel={RSSI (dBm)}]
  \addplot[cGreyD, no marks, line width=0.5pt, densely dashed, forget plot]
    coordinates {(-23,\wlSens) (11,\wlSens)};
  \node[anchor=south east, font=\scriptsize, text=cGreyD, inner sep=1pt]
    at (axis cs:10.5,{\wlSens+2.0}) {MCS7 limit};
  \addplot[cRed, series, mark=*] table[x=tx, y=rssi]
    {\plotdata/wlan_rssi_cpl.dat};
  \addplot[cGrey, series, mark=square*] table[x=tx, y=rssi]
    {\plotdata/wlan_rssi_ant.dat};
  \draw[<->, cBlue, line width=0.8pt]
    (axis cs:\wlGapX,\wlGapLo) -- (axis cs:\wlGapX,\wlGapHi);
  \node[anchor=west, font=\scriptsize\bfseries, text=cBlue, inner sep=2pt]
    at (axis cs:\wlGapX,{0.5*(\wlGapLo+\wlGapHi)}) {\wlGap{} dB};
  \panel[0.04]{b}
\end{axis}
\end{tikzpicture}